\documentclass{article}
\usepackage{spconf,amsmath,graphicx}

\usepackage{makecell}
\usepackage[linesnumbered,ruled,vlined]{algorithm2e}
\usepackage{bm}
\usepackage{amssymb}
\usepackage{multirow}

\def\ie{\emph{i.e.}}
\def\eg{\emph{e.g.}}

\def\etal{\emph{et al.}}
\DeclareMathOperator*{\argmax}{argmax}
\DeclareMathOperator*{\argmin}{argmin}

\SetKwInput{KwInput}{Input}                
\SetKwInput{KwOutput}{Output}              

\begin{document}
\title{SpeechNAS: Towards Better Trade-off between \\ Latency and Accuracy for Large-Scale Speaker Verification}

\name{\begin{tabular}{c}Wentao Zhu*\thanks{*Equal contributions. https://github.com/wentaozhu/speechnas.git}, Tianlong Kong*, Shun Lu, Jixiang Li,\\
Dawei Zhang, Feng Deng, Xiaorui Wang, Sen Yang, Ji Liu\end{tabular}}

\address{Kuaishou Technology}
%
%
%

%

\maketitle

\begin{abstract}
Recently, x-vector~\cite{snyder2018x} has been a successful and popular approach for speaker verification, which employs a time delay neural network (TDNN) and statistics pooling to extract speaker characterizing embedding from variable-length utterances. Improvement upon the x-vector has been an active research area, and enormous neural networks have been elaborately designed based on the x-vector,~\eg, extended TDNN (E-TDNN)~\cite{snyder2019speaker}, factorized TDNN (F-TDNN)~\cite{villalba2019state}, and densely connected TDNN (D-TDNN)~\cite{yu2020densely}. In this work, we try to identify the optimal architectures from a TDNN based search space employing neural architecture search (NAS), named SpeechNAS. Leveraging the recent advances in the speaker recognition, such as high-order statistics pooling, multi-branch mechanism, D-TDNN and angular additive margin softmax (AAM) loss with a minimum hyper-spherical energy (MHE), SpeechNAS automatically discovers five network architectures, from SpeechNAS-1 to SpeechNAS-5, of various numbers of parameters and GFLOPs on the large-scale text-independent speaker recognition dataset {\fontfamily{qcr}\selectfont VoxCeleb1}. Our derived best neural network achieves an equal error rate (EER) of 1.02\% on the standard test set of {\fontfamily{qcr}\selectfont VoxCeleb1}, which surpasses previous TDNN based state-of-the-art approaches by a large margin. 
\end{abstract}
\begin{keywords}
speaker verification, speaker recognition, SpeechNAS, neural architecture search, TDNN
\end{keywords}

\section{Introduction}
\label{sec:intro}
There are numerous measurements and signals, such as fingerprint, face, iris and voice, being investigated for biometric recognition systems~\cite{bimbot2004tutorial}. Among these most popular measurements, voice has been one of the most compelling biometrics, because 1) the microphone system has been one of the most widely adopted intelligent agent to extract the speech signal in various hardwares, and 2) the speech sample can be widely accepted and does not considered threatening by users. Most importantly, the speaker recognition area has been well studied for over fifty years, and there is a rich scientific basis and extensive development over the area.
\begin{figure}[t]
\begin{minipage}[b]{1\linewidth}
  \centering
  \centerline{\includegraphics[trim=30 10 10 10, clip,width=8.5cm]{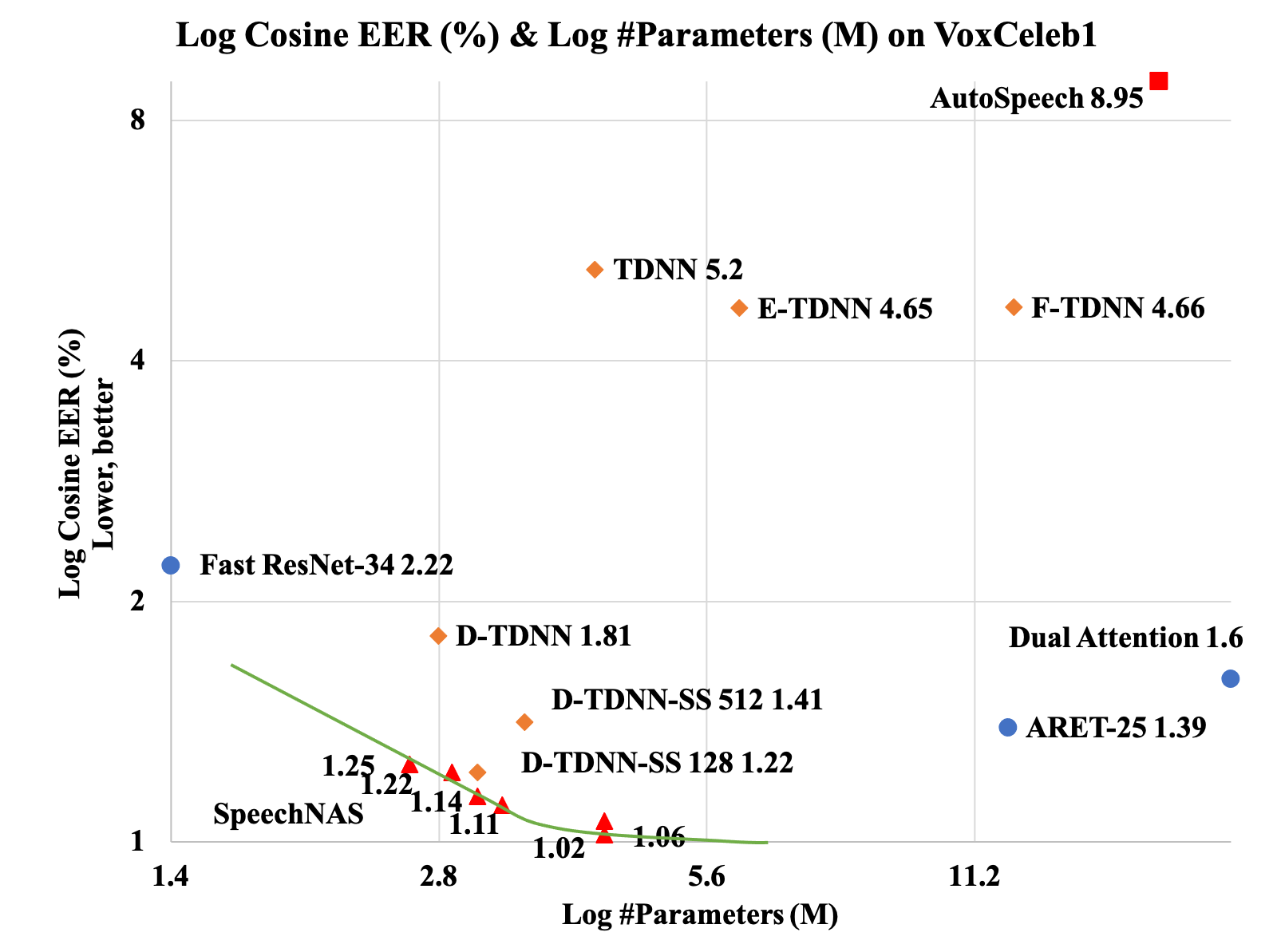}}
\end{minipage}
\caption{SpeechNAS achieves better cosine EERs with lower number of parameters.}
\label{fig:performance_overview}
\end{figure}

Deep neural networks have been widely adopted to extract speaker representations~\cite{snyder2018x}. Recently, time delay neural network (TDNN) with x-vectors~\cite{snyder2018x} has been a paradigm for speaker verification. Compared with vanilla TDNN, extended TDNN (E-TDNN)~\cite{snyder2019speaker} interleaves feed-forward neural network (FNN) layers between the TDNN layer for multi-speaker conversations. Factorized TDNN (F-TDNN)~\cite{villalba2019state} further reduces the number of parameters of the TDNN x-vector by factorizing the weight matrix of each TDNN layer into the product of two low-rank matrices. Densely connected TDNN (D-TDNN)~\cite{yu2020densely} adopts bottleneck layers and dense connectivity, and it can further integrate channel-wise selection mechanism to achieve the state-of-the-art accuracy for TDNN-based speaker verification. The accuracy of TDNN based speaker verification can be further improved leveraging the advanced neural architecture search (NAS).  

On the other hand, studies have been extensively conducted to enhance the loss functions~\cite{liu2019large}. Triplet loss~\cite{zhang2017end} selects appropriate training samples and performs well for speaker verification. Speaker identity subspace loss~\cite{ji2018end} learns a space where it measures the similarity of speakers by Euclidean distance. Center loss~\cite{li2018deep,yadav2018learning} attempts to reduce the intra-speaker variability by constraining features close to the center. Instead, angular distance~\cite{zhang2019utd} focuses on cosine similarity and normalizes the features and the weights of the output layer before softmax. Generalized end-to-end loss~\cite{wan2018generalized} minimizes a scaled cosine score between the features and the estimated speaker centers. Various angular and large margin based loss functions~\cite{huang2018angular,bhattacharya2019adapting,xie2019utterance,li2019boundary} have been investigated in the speaker verification. Liu~\etal~\cite{liu2019large} conducts a comprehensive investigation on the loss functions and combines an additive margin softmax loss and a minimum hyper-spherical energy (MHE)~\cite{liu2018learning} to achieve a desired accuracy. We also employ an advanced large margin based loss to train the candidate architectures for the large scale speaker verification.

In this work, we attempt to construct optimal TDNN-based architectures leveraging the recent advanced study of network designs and loss functions. Our search space consists of multi-branch mechanism, densely connected TDNN (D-TDNN), and channel selection. We conduct the number of branches search, the dimension number of D-TDNN search and the dimension number of channel selection search employing Bayesian optimization. The candidate architecture is trained by a hybrid loss between AAM and a MHE criterion.

Our contributions can be summarized as follows: 1) We design a neural architecture search (NAS) based large scale speaker verification system, SpeechNAS, to identify the optimal architectures leveraging TDNN variants and an AAM related hybrid loss. 2) We conduct a comprehensive comparison to the state-of-the-art speaker verification approaches considering a variety of metrics, including the number of parameters, GFLOPs, latency, the equal error rate (EER)~\cite{snyder2018x}, and the minimum of detection cost function (DCF)~\cite{snyder2018x} with target probabilities set to 0.01 and 0.001. 3) Our NAS based speaker verification derives five architectures, from SpeechNAS-1 to SpeechNAS-5, with various numbers of parameters and FLOPs. The SpeechNAS-5 achieves much better performance than previous TDNN based state-of-the-art approaches on the {\fontfamily{qcr}\selectfont VoxCeleb1} test set as shown in Fig.~\ref{fig:performance_overview}.


\section{Related Works}\label{sec:related}
The recent deep learning based speaker verification approaches can be primarily categorized into two main aspects: advanced network structure constructions~\cite{snyder2018x,snyder2019speaker,villalba2019state,yu2020densely,desplanques2020ecapa} and effective loss function designs~\cite{liu2019large,peng2019logistic,wang2019joint,dhamyal2019optimizing}.

Recently, enormous neural nets have been elaborately designed for speaker recognition, such as TDNN~\cite{snyder2018x}, E-TDNN~\cite{snyder2019speaker}, F-TDNN~\cite{villalba2019state}, D-TDNN~\cite{yu2020densely}. ECAPA-TDNN~\cite{desplanques2020ecapa} employs Res2Net~\cite{gao2019res2net}, SE block~\cite{hu2018squeeze} and channel-dependent attention, which outperforms TDNN based methods. Zhou~\etal~\cite{zhou2019cnn} integrates the phonetic information into the attention based ResNet and improves the speaker verification accuracy significantly. Based on the syllables obtained from the pre-trained HMM models, the SCL~\cite{peng2019syllable} directly improves the discriminative power of the learned frame-level features during training stage. DNN-SAT~\cite{rownicka2019embeddings} investigates the use of embeddings for speaker-adaptive training with a small amount of adaptation data per speaker. LSTM~\cite{zhu2016co,huang2020cycle} can be employed and TDNN-LSTM~\cite{huang2019exploring,zhu2016co} trained with four different data augmentation methods outperforms the baselines of both i-vector~\cite{dehak2010front} and x-vector~\cite{snyder2018x}.  

Various loss functions have been studied for speaker verification. Wang~\etal~\cite{wang2019joint} jointly optimizes classification and clustering with a large margin softmax loss and a large margin Gaussian mixture loss. Logistic affinity loss~\cite{peng2019logistic} instead optimizes an end-to-end speaker verification model by building a learnable decision boundary to distinguish the similar pairs and dissimilar pairs. The quartet loss~\cite{dhamyal2019optimizing} explicitly computes a pair-wise distance loss in the embedding space and increases the gap between the similarity score distributions between the same class pairs and different class pairs. Self-adaptive soft voice activity detection (VAD)~\cite{jung2019self} incorporates a deep neural network based VAD into a deep speaker embedding system to reduce the domain mismatch. Jung~\etal~\cite{jung2019short} applies a teacher-student learning framework to short utterance compensation. 

There are few works for NAS based speaker verification. AutoSpeech~\cite{ding2020autospeech} identifies the optimal operation combination in a neural cell and then derives a ConvNet by stacking the neural cell for multiple times. Auto-Vector~\cite{qu2020evolutionary} searches various choices of temporal context windows based on the x-vector and validates the performance on a private dataset. Concurrent to our work, EfficientTDNN~\cite{wang2021efficienttdnn} searches different depths, kernels, and widths by progressive once-for-all strategy~\cite{cai2019once} for speaker recognition in the wild. We construct the search space based on multi-branch, advanced D-TDNN block and channel-wise selection, and employ Bayesian optimization in the search.  
\section{Method}\label{sec:method}
\begin{figure}[t]
\begin{minipage}[b]{1.0\linewidth}
  \centering
  \centerline{\includegraphics[width=8.5cm]{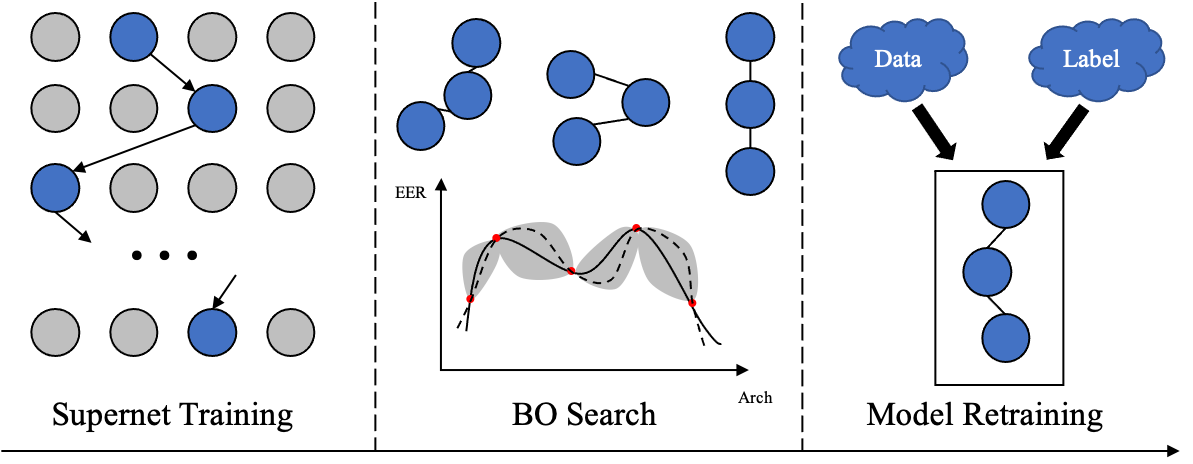}}
\end{minipage}
\caption{The overview of SpeechNAS workflow.}
\label{fig:framework}
\end{figure}

The overall framework of SpeechNAS is illustrated in Fig~\ref{fig:framework}, which consists of supernet construction and training, Bayesian optimization search and candidate architecture retraining.

\subsection{Search Space}\label{sec:search_space}
Let $\mathcal{A}$ be an NAS search space represented by a directed acyclic graph (DAG), and a sub-graph $a \in \mathcal{A}$ is a network architecture denoted as $\mathcal{N}(a, w)$ which is parameterized by weights $w$. The weight sharing strategy~\cite{guo2020single} encodes the search space $\mathcal{A}$ in a supernet $\mathcal{N}(\mathcal{A}, W)$, and all the candidate architectures share the weights $W$ of the supernet. Differentiable NAS~\cite{liu2018darts} further relaxes the discrete search space $\mathcal{A}$ to a continuous one $\mathcal{A}(\theta)$ and jointly optimizes the shared weights $W$ and architecture distribution parameter $\theta$. One shot NAS~\cite{guo2020single} decouples the supernet training and architecture search, which yields a better accuracy. We also conduct two sequential steps for supernet training and architecture search as
\begin{equation}
\begin{aligned}
    W_{\mathcal{A}} &= {\argmin}_{W} \mathcal{L}_{\text{train}}(\mathcal{N}(\mathcal{A}, W)), \\
    a^{\star} &= {\argmin}_{a \in \mathcal{A}} \text{EER}_{\text{val}}(\mathcal{N}(a, W_{\mathcal{A}(a)})),
\end{aligned}
\end{equation}
where $W_{\mathcal{A}}$ is the shared weights after supernet training and $a^{\star}$ is the searched optimal architecture with the best EER on the validation set.

\begin{figure}[t]
\begin{minipage}[b]{1.0\linewidth}
  \centering
  \centerline{\includegraphics[width=8.5cm]{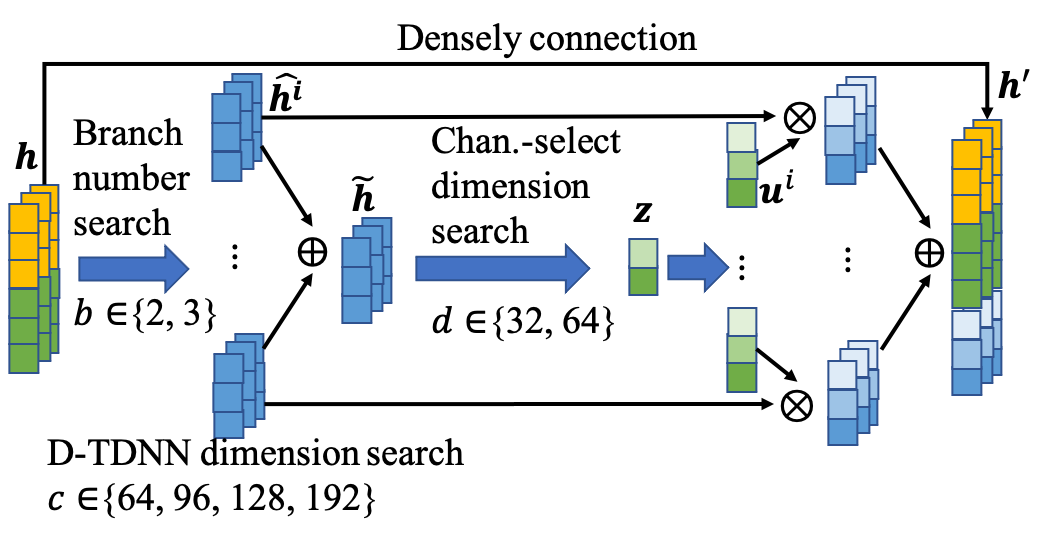}}
\end{minipage}
\caption{The illustration of the SpeechNAS search space.}
\label{fig:search_space}
\end{figure}

Inspired by the superior performance of D-TDNN~\cite{yu2020densely}, we construct the supernet based on D-TDNN blocks, and define the search space $\mathcal{A}$ consisting of the number of branches, the feature dimension of each D-TDNN block, and the dimension of channel-wise selection as illustrated in Fig.~\ref{fig:search_space}.

Let $b \in \mathcal{B}$ be the number of branches where $\mathcal{B}$ is the search branch numbers. We conduct the D-TDNN dimension number search and use $\text{TDNN}_i(\cdot)$ to obtain the $i$-th branch feature for the $t$-th frame $\hat{\bm{h}}^i_t$. Then we search the channel selection dimension numbers and construct a feed-forward net (FNN) layer $\bm{f}$. We obtain the D-TDNN feature $\bm{z}$
\begin{equation}
    \begin{aligned}
    \Tilde{\bm{h}}_t &= \sum_{i=1}^{b} \hat{\bm{h}}^i_t = \sum_{i=1}^{b} (\text{TDNN}_i(\bm{h}))_t, \quad \bm{\mu} = \frac{1}{T} \sum_{t=1}^{T} \Tilde{\bm{h}}_t, \\
    \bm{\sigma} &= \sqrt{\frac{1}{T} \sum_{t=1}^{T} \Tilde{\bm{h}}_t \odot \Tilde{\bm{h}}_t - \bm{\mu} \odot \bm{\mu}}, \quad \bm{s} = \frac{1}{T}\sum_{t=1}^{T} \left(\frac{\Tilde{\bm{h}}_t - \bm{\mu}}{\bm{\sigma}}\right)^3, \\
    \bm{k} &= \frac{1}{T}\sum_{t=1}^{T} \left(\frac{\Tilde{\bm{h}}_t - \bm{\mu}}{\bm{\sigma}}\right)^4, \quad \bm{z} = \bm{f}([\bm{\mu}, \bm{\sigma}, \bm{s}, \bm{k}]),
    \end{aligned}
\end{equation}
where $\bm{h}$ is the feature from the last block and $T$ is the total number of frames. Then we construct a FNN layer $\bm{f}_i^{\prime}$ for the $i$-th branch. The feature $\bm{h}_t^{\prime}$ can be derived
\begin{equation}
    \begin{aligned}
    \bm{t}^i &= \bm{f}_i^{\prime}(\bm{z}), \quad \bm{u}^i = \text{Softmax}(\bm{t}^i), \quad \bm{h}_t^{\prime} = [\bm{h}_t, \sum_{i=1}^{b} \bm{u}^i \odot \hat{\bm{h}}_t^{i} ]
    \end{aligned}
\end{equation}
where the $\text{Softmax}(\cdot)$ is conducted along the branch dimension and $[\cdot, \cdot]$ is the concatenation of the two features.

Specifically, we conduct the number of branches $b$ search and define two options, $2$ or $3$ branches. We expect the branches to learn varied features, and define the dilation rate~\cite{yu2017dilated} of $(1, 3)$, $(1, 3, 5)$ for the two branch options. 
Then we conduct D-TDNN feature dimension number $c$ search and define four options of $64$, $96$, $128$ and $192$. Lastly, we also employ the channel-wise selection based on statistics-and-selection to enhance the feature representational power of SpeechNAS, and conduct the channel selection feature dimension number $d$ search. We define two options of $32$ and $64$ for channel selection feature number search. For each block, there are $2 \times 4 \times 2 = 16$ different candidates. We stack the component into $18$ layers to construct the base feature extractor. The search space consists of $16^{18}$ different candidate architectures, which requires an efficient supernet training strategy and an advanced search algorithm to find optimal architectures.  

\subsection{Loss Function}\label{sec:loss_function}
The large search space leads to slow training of the supernet. To accelerate the training speed of supernet, we employ two different loss functions for the supernet training and the candidate network retraining, respectively. For supernet training, we employ cross entropy loss directly as a proxy loss function to accelerate the supernet training
\begin{equation}
    \mathcal{L}_{\text{train}} = -\frac{1}{N} \sum_{i=1}^{N} \log \frac{e^{\bm{w}_{y_i}^{T}\bm{g}_i}}{\sum_{j=1}^{C}e^{\bm{w}_{j}^{T}\bm{g}_i}},\label{eq:supernet_train}
\end{equation}
where $N$ is the batch size in the stochastic gradient descent (SGD), $y_i$ is the ground truth speaker for the sample, $\bm{w}_{y_i}$ is the related weights for speaker $y_i$ in the last linear layer, $\bm{g}_i$ is the feature vector, and $C$ is the total number of speakers. For the supernet training, we utilize single path $a$ and uniform sampling $U(\mathcal{A})$ strategy to reduce the co-adaptation between node weights~\cite{guo2020single} as
\begin{equation}
    W_{\mathcal{A}} = {\argmin}_{W} \mathbb{E}_{a \sim U(\mathcal{A})} [ \mathcal{L}_{\text{train}}(\mathcal{N}(a, W(a)))].\label{eq:spuniform}
\end{equation}

For the candidate architecture retraining, we employ an additive angular margin softmax (AAM) loss with a minimum hyper-spherical energy (MHE) criterion inspired by the comprehensive investigation of loss functions in the speaker verification~\cite{liu2019large}.  
\begin{gather}
    \begin{aligned}
    \mathcal{L}_{\text{retr}} = &-\frac{1}{N} \sum_{i=1}^{N} \log \frac{e^{s \cdot \cos(\theta_{y_i} + m)}}{e^{s \cdot \cos(\theta_{y_i} + m)} + \sum_{j=1, j\neq y_i}^{C} e^{s \cdot \cos \theta_j} } \\
    &+ \frac{\lambda}{N(C-1)}\sum_{i=1}^{N} \sum_{j=1, j\neq y_i}^{C}\frac{1}{\| \Tilde{\bm{w}}_{y_i} - \Tilde{\bm{w}}_{j} \|^2},
    \end{aligned}\label{eq:retrain}
\end{gather}
where $s$ is a scale factor, $\theta_j$ is the angle between $\bm{w}_j$ and $\bm{g}_i$, $m$ is the additive angular margin to enhance the discriminative power and robustness of feature extractor, $\lambda$ is the trade-off between the AAM loss and MHE regularization, $\Tilde{\bm{w}}_{j}$ is the $L_2$ normalized weights. MHE loss enlarges the inter-class feature separability.
\subsection{Search Algorithm}\label{sec:search_algorithm}
Bayesian optimization (BO)~\cite{shahriari2015taking} iterates between fitting probabilistic surrogate models and determining which configuration to evaluate next by maximizing an acquisition function. We employ a random exploration in the initialization. Gaussian process (GP) with a Hamming kernel $k$ is utilized as the surrogate function, and the generative model of Gaussian process can be defined as
\begin{equation}
    \bm{p} | \bm{a} \sim \mathcal{N}(\bm{\mu}, \bm{K}), \quad \textbf{EER} | \bm{p}, {\sigma}^2 \sim \mathcal{N}(\bm{p}, \sigma^2 \bm{I}),\label{eq:gp}
\end{equation}
where variables $\bm{p}$ are jointly Gaussian, $\bm{a}$ are a set of observed architectures, $K_{ij} = k(a_i, a_j)$, and $\textbf{EER}$ are the evaluated EER values for these architectures with weight sharing. The parameters of GP, $\bm{\mu}$ and $\sigma$, can be estimated by maximizing the marginal log-likelihood. For the acquisition function $\alpha$, we use probability of feasibility (PoF)~\cite{gardner2014bayesian}. 

The overall algorithm is described in Algorithm~\ref{al:speechnas}, which consists of supernet training, architecture search and candidate architecture retraining.
\begin{algorithm}[t]
\SetAlgoLined
\KwInput{Dataset $\mathcal{D}=\mathcal{D}_{train} \cup \mathcal{D}_{val} =\{(\mathbf{x}_i, y_i)| i = 1, \cdots, n\}$, search space $\mathcal{A}$, the number of epochs $n_1$ and the number of candidates $n_2$ in BO, and hyper-parameters in the training}
\KwOutput{Optimal architectures $\bm{a}^\star$ of low EERs on the validation set}
  \tcc{Supernet training}
 Construct a supernet based on the search space $\mathcal{A}$\;
 Train the supernet using equation~\eqref{eq:spuniform} and the loss function in equation~\eqref{eq:supernet_train} with SGD\;
 \tcc{Architecture search}
 Randomly explore $n_2$ candidates $\bm{a}_0$ \tcp*{Initial}
 Evaluate EERs for $\bm{a}_0$ with weight sharing\;
 Add $\bm{a}_0$ and EERs into queue $Q$\;
 Learn GP based on equation~\eqref{eq:gp}\;
 \For{i = 1, 2, $\cdots$, $n_1$}{
  Select new architectures $\bm{a}_i$\ by optimizing acquisition function $\alpha$ \newline $\bm{a}_i = \argmax_{\bm{a}} \alpha(\bm{a}_{i-1}; Q) $\;
  Evaluate EERs for $\bm{a}_i$\;
  Append $\bm{a}_i$ and the EERs to queue $Q$\;
  Update GP based on equation~\eqref{eq:gp} using data in $Q$\;
 }
 \tcc{Candidate networks retraining}
 \For{$a$ in $\bm{a}_{n_1}$}{
 Train network $a$ with SGD and equation~\eqref{eq:retrain}\;
 Save the best trained model and evaluate the EER\;
 }
 \Return Optimal architectures $\bm{a}^\star$\ of low EERs and trained models;
 \caption{The SpeechNAS algorithm}\label{al:speechnas}
\end{algorithm}

\section{Experiment}\label{sec:exper}
\begin{table}[t]
  \caption{The structure of searched optimal networks.}
  \centering
  \begin{tabular}{l|ll|ll|lll}
  \Xhline{2\arrayrulewidth}
    \multirow{2}{*}{Layer} & \multicolumn{2}{c|}{SpeechNAS-3} & \multicolumn{2}{c|}{SpeechNAS-4} & \multicolumn{3}{c}{SpeechNAS-5} \\
    \cline{2-8}
    &$c$ &$d$ &$b$ &$c$ &$b$ &$c$ &$d$ \\
    \Xhline{1.5\arrayrulewidth}
    1 & 96 & 64 &3 &128 & 3 & 128 & 64 \\
    2 & 128 & 32 &3 &64 & 3 & 192 & 32 \\
    3 & 96 & 32 &3 &128 & 3 & 192 & 64 \\
    4 & 96 & 32 &3 &192 & 3 & 192 & 64 \\ \hline
    5 & 96 & 32 &2 &192 & 2 & 192 & 32 \\
    6 & 64 & 64 &2 &64 & 2 & 192 & 64 \\
    7 & 64 & 64 &3 &64 & 3 & 192 & 32 \\
    8 & 96 & 64 &2 &64 & 3 & 128 & 32 \\
    9 & 64 & 64 &3 &64 & 3 & 192 & 64 \\
    10 & 64 & 32 &3 &64 & 2 & 128 & 64 \\
    11 & 64 & 32 &3 &128 & 3 & 128 & 64 \\
    12 & 96 & 64 &3 &192 & 2 & 192 & 64 \\
    13 & 96 & 64 &3 &64 & 2 & 192 & 32 \\
    14 & 96 & 32 &2 &192 & 2 & 128 & 64 \\
    15 & 128 & 32 &2 &128 & 2 & 192 & 32 \\
    16 & 128 & 32 &3 &128 & 2 & 192 & 64 \\
    17 & 96 & 64 &2 &128 & 3 & 192 & 32 \\
    18 & 96 & 64 &3 &192 & 2 & 192 & 64 \\
    \Xhline{2\arrayrulewidth}
  \end{tabular}\label{tab:network}
\end{table}
\begin{table*}[t]
  \caption{Comparisons to state-of-the-art approaches on the {\fontfamily{qcr}\selectfont VoxCeleb1} test set. The \textbf{bold} font denotes the best result. $\star$ denotes training using augmented training set. Latency is measured by averaging the inference time running 1,000 times on one NVIDIA GEFORCE RTX 2080 Ti graphics card with batch size of 128. $\ast$ denotes that batch size of 24 is used for TDNN because of GPU memory size limitation.}  
  \centering
  \begin{tabular}{llllllll}
  \Xhline{2\arrayrulewidth}
    Model & \begin{tabular}{@{}l@{}}Embedding\\ Size\end{tabular} & Parameters (M) & GFLOPs & Latency (ms) & \begin{tabular}{@{}l@{}}Cosine\\ EER (\%)\end{tabular} & DCF\textsubscript{0.01} & DCF\textsubscript{0.001} \\
    \Xhline{1.5\arrayrulewidth}
    Dual Attention~\cite{li2020text} & 512 & 21.7 & - & - & 1.60 & -&- \\ 
    ARET-25~\cite{zhang2020aret} & 512 & 12.2 & 2.9 & - & 1.39 & 0.20 & - \\ 
    Fast ResNet-34~\cite{chung2020defence} & -& 1.4 & 0.45 & - & 2.22 & -  &-\\ 
    TDNN~\cite{snyder2018x} & 512 & 4.2 & 5.34 & 146$\ast$ & 5.20 & 0.44 & 0.60 \\ 
    E-TDNN~\cite{snyder2019speaker} & 512 & 6.1 & 0.91 & 52 & 4.65 & 0.43 & 0.53 \\ 
    F-TDNN~\cite{villalba2019state} & 512 & 12.4 & 2.29 & 115 & 4.66 & 0.41 & 0.57 \\ 
    D-TDNN~\cite{yu2020densely} & 512 & 2.8 & - & - & 1.81 & 0.20 & 0.28 \\ 
    D-TDNN-SS~\cite{yu2020densely} & 512 & 3.5 & 0.56 & 71 & 1.41 & 0.19 & 0.24 \\ 
    D-TDNN-SS~\cite{yu2020densely} & 128 & 3.1 & 0.55 & 70 & 1.22 & 0.13 & 0.20 \\
    \Xhline{1.5\arrayrulewidth}
    AutoSpeech~\cite{ding2020autospeech} & 2048 & 18& -& - & 8.95 & -&- \\ 
    \Xhline{1.5\arrayrulewidth}
    Space (1): SpeechNAS-1 & 128 & 2.6 & 0.44 & 66 & 1.25 & 0.07 & 0.18 \\ 
    Space (1): SpeechNAS-2 & 128 & 2.9 & 0.49 & 68 & 1.22 & 0.07 & 0.17 \\ 
    Space (1): SpeechNAS-3 & 128 & 3.1 & 0.44 & 66 & 1.14 & 0.06 & 0.17 \\ \hline
    Space (2): SpeechNAS-4 & 128 & 3.3 & 0.60 & 62 & 1.11 & 0.06 & 0.24 \\ \hline
    Space (3): SpeechNAS-5 & 128 & 4.3 & 0.76 & 71 & \textbf{1.06} & \textbf{0.06} & \textbf{0.12} \\
    Space (3): SpeechNAS-5$\star$ & 128 & 4.3 & 0.77 & 72 & \textbf{1.02} & \textbf{0.05} & \textbf{0.17} \\
    \Xhline{2\arrayrulewidth}
  \end{tabular}\label{tab:sota}
\end{table*}
\begin{table}[t]
  \caption{Cosine EERs (\%) of loss functions for supernets.}
  \centering
  \begin{tabular}{llll}
  \Xhline{2\arrayrulewidth}
    Model & Searched arch.s & Retrain \\
    \Xhline{1.5\arrayrulewidth}
    Space (2) w/ cross entropy & 1.92 - 2.06 & 1.11 \\ 
    Space (2) w/ AAM + MHE & 1.54 - 1.61 & 1.13 \\ \hline
    Space (3) w/ cross entropy & 1.72 - 1.80 & 1.06 \\
    Space (3) w/ AAM + MHE & 1.29 - 1.35 & 1.09 \\
    \Xhline{2\arrayrulewidth}
  \end{tabular}\label{tab:loss}
\end{table}
\subsection{Dataset}\label{sec:exper_dataset}
\textbf{Standard training set} \, For the standard training without augmentation, we follow the same dataset preparation in D-TDNN~\cite{yu2020densely} for {\fontfamily{qcr}\selectfont VoxCeleb1}~\cite{nagrani2017voxceleb} and {\fontfamily{qcr}\selectfont VoxCeleb2}~\cite{chung2018voxceleb2}. The two versions have their own explicit train and test splits and consist of 7,323 speakers with over one million utterances and more than 2,000 hours audio data. The training samples are generated by following the Kaldi toolkit~\cite{povey2011kaldi} recipe. We use the whole {\fontfamily{qcr}\selectfont VoxCeleb2} and the training set of {\fontfamily{qcr}\selectfont VoxCeleb1} as our training set, and validate our method on the test set of {\fontfamily{qcr}\selectfont VoxCeleb1}.

For the preprocessing, we extract 30-dimensional Mel-frequency cepstrum coefficients (MFCCs) over a 25 ms long window every 10 ms. To remove silent frames, cepstral mean normalization (CMN) is applied over a 3 seconds long sliding window and energy based VAD is utilized. We randomly split the spectrograms into 200 to 400 frames.

\textbf{Generalization on augmented training set} \, We further validate the \emph{generalization ability} of our explored optimal network on an augmented dataset. We only augment the training set of {\fontfamily{qcr}\selectfont VoxCeleb2} and follow the same procedure of ECAPA-TDNN~\cite{desplanques2020ecapa} for the data augmentation, which generates a total of six extra samples for each utterance. We utilize the Kaldi recipe~\cite{povey2011kaldi} in combination with the publicly available MUSAN dataset (babble and noise)~\cite{snyder2015musan} and the RIR dataset (reverb)~\cite{ko2017study} for the first three augmentations. We generate the remaining three augmentations using the open-source SoX~\cite{barras2012sox} (tempo up and tempo down) and FFmpeg (alternating opus or aac compression) libraries.  

We extract 80-dimensional MFCCs from a 25 ms window with a 10 ms frame shift as the input features on the augmented training set. We utilize the CMN to normalize two second random crops of the MFCCs feature vectors. We apply SpecAugment~\cite{park2019specaugment} on the log Mel spectrogram of the samples for the augmentation. In the frequency domain, we randomly mask zero to five frames in the time domain and zero to ten channels in the frequency domain.

\subsection{Implementation Detail}\label{sec:exper_impl}
We implement the whole framework based on PyTorch. For the training, we use SGD with momentum of 0.95 and the initial learning rate of 0.01. The weight decay is set to $5\times 10^{-4}$, and batch size is set to 128. We use 26 epochs for both the supernet training and candidate network retraining. The learning rate is adapted to 0.001 at the epoch of 14 and 0.0001 at the epoch of 20.  

The scaling factor $s$ in the AAM loss is set to be 30 and the margin $m$ is set to be 0.2. The strength of the MHE criterion $\lambda$ is set to be 0.01. These hyper-parameters are set basically according to those in D-TDNN~\cite{yu2020densely} firstly and tuned a little due to the variant in our method and re-implementation. We use the reported performance directly from published papers for previous state-of-the-art approaches in Table~\ref{tab:sota}. We use the public implementations\footnote{https://github.com/cvqluu/TDNN; https://github.com/cvqluu/Factorized-TDNN} to calculate the latency, GFLOPs and the number of parameters (M) of TDNN and F-TDNN. We implement E-TDNN and D-TDNN, and reproduce the results of E-TDNN and D-TDNN. For D-TDNN-SS, we achieve an EER of 1.24\% compared to reported 1.22\%.

In the search phase, we set $n_1$ to be 100 and $n_2$ to be 64. To train a well-performed Gaussian process model, we randomly evaluate 1,200 architectures in the initialization step. For both the supernet training and candidate architecture retraining, we use one single NVIDIA GEFORCE RTX 2080 Ti graphics card. We use eight 2080 Ti graphics cards to accelerate the search phase. We utilize the OpenBox library~\cite{li2021openbox} to implement the Bayesian optimization in the search. 

\subsection{Result}\label{sec:exper_result}
We conduct a comprehensive investigation of our SpeechNAS and adopt six metrics for evaluation, including the number of parameters (M), GFLOPs, Latency (ms), the EER with cosine similarity scoring~\cite{snyder2019speaker} and the minimum of detection cost function (DCF)~\cite{snyder2019speaker} with target probabilities set to 0.01 and 0.001. 

The structure of searched optimal networks are shown in Table~\ref{tab:network}. From Table~\ref{tab:network}, we observe that the first four layers of all the three searched networks have large numbers of parameters, probably because the large first few layers encourage various feature exploration. Our SpeechNAS automatically design complicated network structures which are difficult to be manually designed.  

\textbf{The effect of search spaces} \, We try three different search space design strategies, \textbf{(1)} both the D-TDNN feature and channel selection dimension number search \{\{2\}, \{64, 96, 128\}, \{32, 64\}\},~\ie, without the number of branches search, \textbf{(2)} the number of branches search and the D-TDNN feature dimension number search \{\{2, 3\}, \{64, 128, 192\}, \{32\}\},~\ie, without the channel selection feature dimension number search, \textbf{(3)} the full search space \{\{2, 3\}, \{128, 192\}, \{32, 64\}\},~\ie, the number of branches search and the feature dimension number search for each D-TDNN block and channel selection. For \textbf{search space (1)}, we obtain three optimal candidate architectures with various numbers of parameters, GFLOPs, latency and EERs, named SpeechNAS-1, SpeechNAS-2, and SpeechNAS-3. For \textbf{search space (2)}, we obtain SpeechNAS-4. For the full \textbf{search space (3)}, we obtain SpeeechNAS-5. The comprehensive performance comparison is listed in Table~\ref{tab:sota}.  

From Table~\ref{tab:sota}, SpeechNAS-5 yields a better EER than SpeechNAS-4 and SpeechNAS-3, which shows that the full \textbf{search space (3)} offers the best EER. SpeechNAS-4 yields a better EER than SpeechNAS-3, which demonstrates the accuracy gain from search on branch numbers is larger than that from DTDNN feature dimensions in the SpeechNAS. For the model complexity, the number of parameters in SpeechNAS-4 is a little larger than SpeechNAS-3, and the number of parameters in SpeechNAS-5 is the largest. Although the full \textbf{search space (3)} offers the best EER, the model complexity of obtained architecture SpeechNAS-5 is the largest. From the latency perspective, the SpeechNAS-5 only increases a little compared to SpeechNAS-3.

\textbf{The effect of loss functions} \, For the supernet training, we investigate using cross entropy in equation~\eqref{eq:supernet_train} and AAM with MHE in equation~\eqref{eq:retrain} in Table~\ref{tab:loss}. Because the AAM with MHE induces a better EER, the EERs of searched architectures using AAM + MHE without retraining are better than supernet trained with cross entropy. However, the retraining cannot yield better EERs for supernet training with AAM + MHE, probably because cross entropy loss enables a more adequate training for weights in the supernet which leads to a more accurate performance estimator than AAM + MHE in the weight sharing. Because training time of supernet with AAM + MHE typically is more than twice than that of cross entropy loss, we use cross entropy loss to train the supernet.

\textbf{Comparison to state-of-the-art approaches} \, We conduct a comprehensive comparison of SpeechNAS to previous TDNN based state-of-the-art methods in Table~\ref{tab:sota}. From Table~\ref{tab:sota} and Fig.~\ref{fig:performance_overview}, our SpeechNAS achieves much better accuracy with a low latency which is friendly to be deployed to data intensive applications. Retraining the SpeechNAS-5 on an augmented dataset yields further EER improvement, which demonstrates a good generalization of SpeechNAS. We notice that ECAPA-TDNN~\cite{desplanques2020ecapa} (embedding size 512) of 6.2M parameters, which is based on Res2Net~\cite{gao2019res2net} and SE-ResNet~\cite{hu2018squeeze}, achieves 1.01\% EER. Our SpeechNAS is based on TDNN framework and achieves a comparable accuracy with much less number of parameters of 4.3M.

\section{Conclusion}\label{sec:conclu}
In this work, we introduce a new neural architecture search based speaker verification framework, called SpeechNAS. Specifically, we investigate the optimal TDNN based network architectures leveraging the recent advances of the multiple branches design, D-TDNN block, channel-wise selection and the additive angular margin softmax loss (AAM) with a minimum hyper-spherical energy (MHE) criterion in the signal processing. We construct a supernet based on the designed search space and employ single path and uniform sampling with cross entropy loss to efficiently train the supernet. For the search, we utilize the Bayesian optimization with Gaussian process to find the best performed candidate architectures. The candidate architectures are retrained with a hybrid loss of AAM and MHE. Ablation studies and comprehensive experimental results on a large-scale speaker verification dataset, {\fontfamily{qcr}\selectfont VoxCeleb1}, demonstrate the effectiveness of each component of our SpeechNAS and that our SpeechNAS achieves a much better accuracy than other TDNN based state-of-the-art variants with a reasonable model complexity. Future NAS related work can be conducted to further improve the speaker verification accuracy by designing advanced search spaces with more effective components, blocks and/or attention mechanisms.

{\small
\bibliographystyle{IEEEbib}
\bibliography{strings}  
}

\end{document}